\documentclass[final,5p,times,twocolumn,number]{elsarticle}

\biboptions{square,sort&compress}

\usepackage{amssymb}
\usepackage{amsmath}
\usepackage{mathrsfs}
\usepackage{lipsum}
\usepackage[dvipsnames]{xcolor}
\usepackage{tikz}
\usetikzlibrary{arrows.meta}
\usepackage{pgfplots}
\pgfplotsset{compat=1.18}
\usetikzlibrary{calc}
\usetikzlibrary{3d}

\usepackage[
    colorlinks=true,
    citecolor=ForestGreen,
    linkcolor=black,
    urlcolor=MidnightBlue
]{hyperref}

\journal{Physics Letters B}
\begin{document}
\begin{frontmatter}

    \title{2-Group global symmetry in the compactified M2-brane}

\author[first]{F. Caro P\'erez}
\affiliation[first]{organization={Departamento de Física, Universidad de Antofagasta},%Department and Organization
            addressline={Av. Universidad de Antofagasta}, 
            city={Antofagasta},
            postcode={02800}, 
            state={Antofagasta},
            country={Chile}}
\author[second,third]{M.P García del Moral}
\affiliation[second]{organization={Departamento de Química, Área de Física, Universidad de la Rioja},%Department and Organization
            addressline={
C/ Madre de Dios 53}, 
            city={Logroño},
            postcode={26006}, 
            state={La Rioja},
            country={Spain}}
\affiliation[third]{organization={Instituto de Investigación en Computación Científica (SCRIUR),  Universidad de la Rioja},%Department and Organization
            addressline={C/ Madre de Dios 53}, 
            city={Logroño},
            postcode={26006}, 
            state={La Rioja},
            country={Spain}}
           
\author[cuart]{A. Restuccia}
\affiliation[cuart]{organization={Departamento de Física, Universidad de Antofagasta},%Department and Organization
            addressline={Av. Universidad de Antofagasta}, 
            city={Antofagasta},
            postcode={02800}, 
            state={Antofagasta},
            country={Chile}}
\begin{abstract}
We study generalized global symmetries of the bosonic M2-brane in eleven-dimensional backgrounds with non-trivial four-form flux. Focusing on compactifications of the form
\(AdS_5\times S^2_{1} \times S^2_{2}\times T^2\), we show that the monopole and winding symmetry sectors of the membrane do not organize as an ordinary direct product once the Wess--Zumino coupling is included. Instead, they combine into a non-trivial \(2\)-group global symmetry. We identify the corresponding mixed background-gauge structure and show that the quantized target space flux determines the associated \textit{Postnikov} class. This provides a concrete realization of higher-group symmetry in the worldvolume theory of the M2-brane induces a flux quantization on the worldvolume and relates its global symmetry structure to the introduction of a quantized flux in M-theory.
\end{abstract}

\end{frontmatter}
\tableofcontents

%%%%%%%%%%%%%%%%%%%%%%%%%%%%%%%%%%%%%%%%%%%%%%%%%%%%%%%%%%%%%%%%%%%%%%%%%%%%%%%%%%%%%%%%%%%%%%%%
\section{Introduction}
Symmetry has undergone a substantial conceptual refinement during the last decade. In its conventional formulation, a continuous global symmetry is associated with a conserved Noether current. The charge of this current acts on local operators and thereby organizes the spectrum of states of the theory.  The modern viewpoint enlarges this notion by identifying global symmetries with topological operators. In this formulation, the charged objects need not to be local insertions, but they may instead be line operators, surface operators, or higher-dimensional defects. This leads naturally to the notion of higher-form global symmetries (HFS), which provide the appropriate language for quantum field theories containing extended charged objects \cite{Gaiotto:2014kfa}.

\noindent
This perspective is especially natural in string theory and M-theory \cite{Albertini:2020mdx,Franco:2024mxa,Bharadwaj:2024gpj,Heckman:2024obe}. Their spectra contain extended degrees of freedom, such as strings, membranes, and higher-dimensional
branes, which couple to higher-form gauge potentials. In particular, the M2-brane couples electrically to the eleven-dimensional three-form connection \(C_{[3]}\) through a Wess--Zumino term \cite{Bergshoeff:1987cm}. As a consequence, the worldvolume dynamics of the membrane becomes sensitive not only to the local geometry of the target space, but also to its global topology and to the quantization of background fluxes \cite{Witten:1995ex,Bergshoeff:1987cm,Caro-Perez:2026pbu}.

\noindent
Once generalized global symmetries are identified, an important question concerns the way that symmetry sectors of different form degree combine. In general, for theories containing HFS of degree zero and one, the global symmetry structure need not factorize as a direct product of independent HFS$(0)$ and HFS$(1)$ groups \cite{Benini:2018reh}. Instead, these sectors may be intertwined into a
higher-group structure. The simplest non-trivial possibility is a \(2\)-group global
symmetry, in which an ordinary \(0\)-form symmetry and a \(1\)-form symmetry are combined through a non-trivial extension controlled by a Postnikov class \cite{Sharpe:2015mja,Cordova:2018cvg,Benini:2018reh}.

\noindent
A useful diagnostic of a \(2\)-group symmetry is provided by studing the allowed background gauge fields. Let \(B_{[1]}\) denote a background connection for a HFS$(0)$ and let \(\widetilde B_{[2]}\) denote a two-form background field for a HFS$(1)$. If the symmetry were an ordinary direct product, these backgrounds could be gauged independently. In a genuine \(2\)-group, however, the gauge transformation of \(\widetilde B_{[2]}\) acquires an additional contribution induced by the \(0\)-form gauge parameter. Consequently, the naive curvature \(d\widetilde B_{[2]}\) is replaced
by a modified higher curvature,
\begin{align}
    H_{[3]}={}&d\widetilde B_{[2]}-\Omega_{[3]}(B_{[1]}),
\end{align}
whose exterior derivative obeys a twisted Bianchi identity,
\begin{align}
    dH_{[3]}
    =
    -
    P_{[4]}(dB_{[1]}).
\end{align}
The cohomology class represented by \(P_{[4]}\) is the Postnikov class of the \(2\)-group. It measures the obstruction to factorizing the symmetry as an ordinary
product of a \(0\)-form and a \(1\)-form sector \cite{Benini:2018reh}. Continuous \(2\)-group symmetries arise in four-dimensional quantum field theories through deformed current algebras with quantized structure constants, in which two ordinary flavor currents may fuse into a conserved two-form current \cite{Cordova:2018cvg}. Closely related mechanisms appear in six-dimensional gauge theories and little string theories, where mixed anomaly data can organize ordinary and higher-form symmetries
into non-trivial \(2\)-groups \cite{Cordova:2020tij}. More recently, mixed Wess--Zumino--Witten terms in effective chiral Lagrangians were shown to encode
\(2\)-group global symmetries intertwining a flavor symmetry with a winding HFS$(1)$ \cite{Davighi:2024zjp}. In those examples, the quantized coefficient of the topological
term coincides with the corresponding Postnikov class.

\noindent
Higher-group symmetries also arise naturally in string- and M-theoretic settings \cite{DelZotto:2022joo}. The geometric origin of higher-form symmetries in M-theory can be extracted from defect groups and flux data of the underlying background \cite{Albertini:2020mdx}, \cite{Caro-Perez:2026pbu}, \cite{Caro-Perez:2026wxd}. Moreover, \(2\)-group structures have been identified in quantum field theories engineered from M-theory compactifications,
where the mixing between HFS$(0)$ and HFS$(1)$ backgrounds is encoded in the boundary geometry of the M-theory background \cite{DelZotto:2022joo}. These developments indicate that \(2\)-group symmetry is a natural organizing principle whenever higher-form charges, compactification data, and topological couplings mixes. The purpose of the present work is to show a direct worldvolume realization of such a \(2\)-group structure in the bosonic M2-brane. To illustrate it we consider an eleven-dimensional
background of the form \cite{Wulff:2016vqy,Figueroa-OFarrill:2011tnp}:
\begin{align}
    \mathcal{M}_{11}=AdS_5\times S^2_1 \times S^2_2 \times T^2,
\end{align}
together with a non-trivial four-form supergravity flux with a constant coefficient. The two-torus supports compact membrane embedding coordinates and therefore winding sectors, whereas the two spherical factors and the torus carry the monopole-type two-form flux entering the Wess--Zumino coupling. These ingredients give rise, respectively, to a HFS\((1)\) winding symmetry and to a HFS\((0)\) compact shift symmetry whose current is improved by the monopole potential in the compact sector of the M2-brane worldvolume theory.

\noindent
In the absence of the flux-induced Wess--Zumino coupling, the monopole and winding sectors would organize as an ordinary direct product of generalized global symmetries as described in \cite{Caro-Perez:2026pbu}.
The central observation of this work is that the WZ-term coupling obstructs this factorization. Once one couples the corresponding currents to background gauge fields, the gauge variation of the M2-brane action can no longer be cancelled by independent transformations of the \(0\)-form and \(1\)-form backgrounds. Instead, the two-form background associated with the winding sector must transform non-trivially under the gauge transformation of the monopole-sector background. This is precisely the characteristic signature of a \(2\)-group global symmetry.

%%%%%%%%%%%%%%%%%%%%%%%%%%%%%%%%%%%%%%%%%%%%%%%%%%%%%%%%%%%%%%%%%%%%%%%%%%%%%%%%
\section{The supergravity background with compact sector}
\label{sec:ads5-s2-s2-t2-background}

As explained before we consider a symmetric source-free background, an exact solution of eleven-dimensional supergravity which contains a flat two-torus. 
In this sense we consider the symmetric product background
\begin{align}
    \mathcal{M}_{11}
    =
    \mathrm{AdS}_5(L)
    \times
    S^2_1(R_1)
    \times
    S^2_2(R_2)
    \times
    T^2 .
\end{align}
Here \(L\) denotes the \(\mathrm{AdS}_5\) radius and \((R_1,R_2)\) denotes the radius of the two spherical factors. These radii are fixed by the scale of the four-form flux. In the full M-theory setting, the appropriate dimensionless
parameter is \(L/\ell_p\), and the supergravity approximation requires \(L/\ell_p\gg 1\). This geometry belongs to the standard class of symmetric space solutions of eleven-dimensional supergravity. We use it as the gravitationally consistent setting in which the M2-brane retains a toroidal winding sector. We denote by
\begin{align}
    \omega_1=\omega(S^2_1),
    \qquad
    \omega_2=\omega(S^2_2),
    \qquad
    \omega_T=\omega(T^2)
\end{align}
the volume forms of the two spherical factors and of the torus. They are normalized so that their periods are integral, namely
\begin{align}\label{eq:normalization-volumen-forms-spheres}
    \left[
    \frac{\omega_1}{2\pi}
    \right]\in H^2(S^2_1,\mathbb Z),
    \quad
    \left[
    \frac{\omega_2}{2\pi}
    \right]\in H^2(S^2_2,\mathbb Z),
    \quad
    \left[
    \frac{\omega_T}{2\pi}
    \right]\in H^2(T^2,\mathbb Z).
\end{align}
In particular,
\begin{align}
    d\omega_1=0,
    \qquad
    d\omega_2=0,
    \qquad
    d\omega_T=0.
\end{align}
The four-form flux is chosen as
\begin{align}
    F_{[4]}
    =
    2f_1\,\omega_1\wedge\omega_2
    +
    f_2\,\omega_1\wedge\omega_T
    +
    f_3\,\omega_2\wedge\omega_T .
    \label{eq:F4-ads5-s2-s2-t2}
\end{align}
Where $(f_1,f_2,f_3)\in \mathbb{Z}^3$ are coefficients with a \textit{Algebraic Einstein Constrain} (\textit{AEC}) given by $2
f_1^2=f_2^2+f_3^2$ \cite{Wulff:2016vqy} \textit{i.e.} a real cuadratic cone in the parameter-space. The 4-form is quantized:
\begin{align}\label{eq: flux_quantization}
\int_{\mathscr{C}_4}F_{[4]}=2f_1+f_2+f_3=k\in \mathbb{Z}^\times,  
\end{align}
where we use the notation $\mathbb{Z}^\times\equiv\mathbb{Z}-\{0\}$.
This is the symmetric branch relevant for the present construction. 

\noindent
The flux contains
two mixed components involving the torus,
\begin{align}
    f_2\,\omega_1\wedge\omega_T
    +
    f_3\,\omega_2\wedge\omega_T,
\end{align}
which are the components that couple the toroidal winding sector to the monopole
data of the spherical factors, and the additional component 
\begin{align}
    2f_1\,\omega_1\wedge\omega_2,
\end{align}
required by the eleven-dimensional Einstein equations. It allows the two spheres
to curve consistently while the torus remains flat. This particular choice of parameters do not inpose an additional dynamical assumption.

\noindent
In an orthonormal frame adapted to the product geometry, the flux can be written as
\begin{align}
    F_{[4]}
    =
    2f_1\,e^{5678}
    +
    f_2\,e^{569\,10}
    +
    f_3\,e^{789\,10},
\end{align}
where \(e^5,e^6\) span(\(T^*S^2_1\)), \(e^7,e^8\) span(\(T^*S^2_2\)), and
\(e^9,e^{10}\) span(\(T^*T^2\)) and $e^{ijkl}:=e^i\wedge e^j \wedge e^k \wedge e^l$. The relations among the radii are
\begin{align}
    L^2=\frac{4}{f_1^2},
    \qquad
    R^2_1=\frac{2}{4f_1^2-f_3^2},\qquad R_2^2=\frac{2}{4f_1^2-f_2^2}.
    \label{eq:ads5-s2-radius-relations}
\end{align}
Equivalently,
\begin{align}
    R_{mn}(\mathrm{AdS}_5)
    &=
    -\frac{1}{6}(4f_1^2+f_2^2+f_3^2)\,g_{mn},
    \\
    R_{ab}(S^2_1)
    &=
    \frac{1}{6}(8f_1^2+2f_2^2-f_3^2)g_{ab}^{(1)},
    \\
    R_{cd}(S^2_2)
    &=
    \frac{1}{6}(8f_1^2+2f_3^2-f_2^2)g_{cd}^{(2)},
    \\
    R_{rs}(T^2)
    &=
    0.
\end{align}
Here $m,n$ are indices along $\mathrm{AdS}_5$, $a,b$ along $S^2_1$,
$c,d$ along $S^2_2$, and $r,s$ along the target-space torus $T^2$.
The last equation shows that the torus remains Ricci-flat. This is
essential for the global symmetry description.

\noindent
We use \eqref{eq:F4-ads5-s2-s2-t2} as the starting point for the worldvolume analysis. This background is the appropriate source-free supergravity setting for the \(2\)-group structure studied below. Let \(A_1\) and \(A_2\) be local monopole potentials satisfying
\begin{align}
    dA_1=\omega_1,
    \qquad
    dA_2=\omega_2.
\end{align}
Since \(A_1\) and \(A_2\) are gauge 1-conecction, they are local representatives rather
than globally defined one-forms. This is the usual local description of a non-trivial line bundle over \(S^2\) (the chern class) \cite{Brylinski:1993ab}, associated to the existence of monopoles over the spheres. Thus the ansatz has an extra $\mathbb{Z}_2$ symmetry in the case when  
\begin{align}
    S^2_1\leftrightarrow S^2_2,
    \qquad
    f_2\leftrightarrow f_3,
    \qquad
    \omega_1\leftrightarrow \omega_2.
\end{align}
In the generic case, \(f_2\neq f_3\), the two spherical factors have different
radii, \(R_1\neq R_2\) (related by the same exchange symmetry).

\noindent
A local representative of the supergravity 3-form potential is
\begin{align}
    C_{[3]}
    =
    f_1(\,A_1\wedge\omega_2
    +\,A_2\wedge\omega_1)
    +(f_2\,A_1+f_3\,A_2)\wedge\omega_T.
    \label{eq:C3-ads5-s2-s2-t2}
\end{align}
Indeed it can be straithfawardly verified that $dC_{[3]}=F_{[4]}$. The first two terms in \eqref{eq:C3-ads5-s2-s2-t2},
\begin{align}
    f_1(\,A_1\wedge\omega_2
    +\,A_2\wedge\omega_1)
\end{align}
belong purely to the spherical sector. They are needed for the eleven-dimensional
supergravity flux, but they do not involve the toroidal embedding fields of the
M2-brane. The term relevant for the toroidal winding sector is instead
\begin{align}
    C_{[3]}^{(T^2)}
    =(f_2\,A_1+f_3\,A_2)\wedge\omega_T.
\end{align}
It is useful to define
\begin{align}
    \mathcal A_{[1]}
    :=
    f_2\,A_1+f_3\,A_2,
    ~~\text{with}~~
    \mathcal F_{[2]}
    :=
    d\mathcal A_{[1]}
    =
    (f_2\omega_1+f_3\omega_2).
    \label{eq:calA-calF-def}
\end{align}
The form \(\mathcal A_{[1]}\) is a local monopole potential for the diagonal
monopole class carried by the two spherical factors. Its curvature
\(\mathcal F_{[2]}\) is globally defined and is the geometric datum that will enter the
\(2\)-group extension.

\noindent
Let \(X^r: \Sigma_3\longrightarrow T^2\) with $\Sigma_3$ he wordlvolume of M2-brane and \(r=1,2\), denote the embedding maps from $\Sigma_3$ to \(T^2\).
Then the pullback $\mathbb{P}(.)$ over the forms is defined by
\begin{align}
    \mathbb P(\omega_T)
    =
    q\,dX^1\wedge dX^2
    =
    \frac{q}{2}\,\epsilon_{rs}\,
    dX^r\wedge dX^s.
\end{align}
where $q\in \mathbb{Z}^\times$ \cite{GarciaDelMoral:2018jye}. Therefore, the toroidal part of the Wess--Zumino coupling is
\begin{align}
S_{\mathrm{WZ}}^{(T^2)}
    =
    \int_{\Sigma_3}
    \mathbb P\!\left(
    \mathcal A_{[1]}\wedge\omega_T
    \right)
    =
    \frac{q}{2}
    \int_{\Sigma_3}
    \mathbb P(\mathcal A_{[1]})
    \wedge
    \epsilon_{rs}\,
    dX^r\wedge dX^s.
    \label{eq:WZ-torus-sector}
\end{align}
In what follows we suppress the symbol \(\mathbb P\) on background forms whenever no confusion can arise. Thus \(\mathcal A_{[1]}\), \(\mathcal F_{[2]}\), \(\omega_1\),
\(\omega_2\), and \(\omega_T\) will denote their pullbacks to the M2-brane worldvolume
or to the auxiliary manifold used for inflow.

\section{Global symmetries of M2-brane without WZ-term}
\label{sec:global-symmetries-compact-sector}
Before coupling the compact sector to background gauge fields, let us identify the global symmetries of the ungauged theory. The M2-brane action in \textit{Polyakov-frame} is
\begin{align}
    S_{M2}={}&-\frac{T_{\mathrm{M2}}}{2}\int_{\Sigma_3}
        dX^r\wedge\star dX_r    
    +\frac{q}{2}
    \int_{\Sigma_3}
    \mathcal A_{[1]}\wedge
    \epsilon_{rs}dX^r\wedge dX^s+\notag
    \\
    {}&-\frac{T_{M2}}{2}\int_{\Sigma_3}\omega_{vol}(\Sigma_3) -\frac{T_{M2}}{2}\int_{\Sigma_3}dX^m\wedge \star dX_m-\notag
    \\
    {}&-\frac{T_{M2}}{2}\int_{\Sigma_3}dX^a\wedge \star dX_a+\frac{1}{2}\int_{\Sigma_3}f_1\epsilon^{ab}A_a\wedge \omega_b.
    \label{eq:ungauged-torus-action}
\end{align}
For simplicity, we set the M2-brane tension to \(T_{\mathrm{M2}}=1\) and
\begin{align}
S_{M2}^{nc+S^2}:={}&-\frac{1}{2}\int_{\Sigma_3}\omega_{vol}(\Sigma_3)-\frac{1}{2}\int_{\Sigma_3}(dX^m\wedge \star dX_m+\notag
\\
{}&+dX^a\wedge \star dX_a)+\frac{1}{2}\int_{\Sigma_3}f_1\epsilon^{ab}A_a\wedge \omega_b.
\end{align}
The compact fields \(X^r\), \(r=1,2\), are embedding $$X^r(\xi^1, \xi^2): \Sigma_3\longrightarrow T^2,$$ along the two-torus where $\Sigma_3$ denotes the closed-compact $(2+1)$ M2-brane wordlvolume with topology of $\Sigma_3\cong T^2\times S^1$. Therefore the theory has a compact shift symmetry
\begin{align}\label{eq:26}
    X^r
    \longmapsto
    X^r+\varepsilon^r,
    \qquad
    \varepsilon^r=\mathrm{constant}.
\end{align}
Since the action depends on \(X^r\) only through \(dX^r\), this is a global \(U(1)^{(0)}_m\times U(1)^{(0)}_m\) symmetry \cite{Caro-Perez:2026pbu}.
The equations of motion for the compact fields \(X^r\) follow from varying the
toroidal part of the action $S_{M2}^{c}$\footnote{We denote $S_{M2}^{c}$ as the compact part of \eqref{eq:ungauged-torus-action}.}
\begin{align}
    S_{M2}^{c}={}&
    -\frac12
    \int_{\Sigma_3}
    dX^r\wedge\star dX_r
    +
    \frac q2
    \int_{\Sigma_3}
    \mathcal A_{[1]}\wedge
    \epsilon_{rs}dX^r\wedge dX^s .
    \label{eq:ungauged-torus-action}
\end{align}
Since \(\mathcal A_{[1]}\) is a fixed background one-form on the spherical factors, the
variation is taken only with respect to the toroidal embedding coordinates \(X^r\),
One obtains the equation of motion
\begin{align}\label{eq:eom-ungauged-Xr}
    d\star dX_r+
    q\epsilon_{rs}\mathcal F_{[2]}\wedge dX^s={}&0 .
\end{align}
This equation can be written as the conservation of the modified monopole current
\begin{align}
    J_r
    =
    \star dX_r
    +
    q\epsilon_{rs}\mathcal A_{[1]}\wedge dX^s .
    \label{eq:ungauged-shift-current}
\end{align}
Thus the equations of motion are equivalent to
\begin{align}
    dJ_r=0.
\end{align}
The Wess--Zumino coupling therefore improves the ordinary shift current by a Page-like term involving the local monopole potential \(\mathcal A_{[1]}\) \cite{Marolf:2000cb}. This is
the main difference with respect to the target space discussed in \cite{Caro-Perez:2026pbu}.

\noindent
The other global symmetry is associated with the winding current (HFS$(1)$):
\begin{align}
    j^r_{\mathrm w}
    =
    dX^r .
    \label{eq:winding-current}
\end{align}
This current is conserved identically by the Bianchi identity
\begin{align}
    d j^r_{\mathrm w}={}&0.
\end{align}
It generates a \(U(1)^{(1)}_{\mathrm{w}}\) winding symmetry. Its charged objects are line defects
or winding sectors of the compact M2-brane embedding. Equivalently, the corresponding
topological symmetry operator is supported on closed one-cycles of \(\Sigma_3\) \cite{Caro-Perez:2026pbu}.

\noindent
Thus, before coupling to backgrounds, the compact sector contains two generalized
symmetry sectors:
\begin{align}
    U(1)^{(0)}_{\mathrm{m}}
    \qquad\text{and}\qquad
    U(1)^{(1)}_{\mathrm{w}}.
\end{align}
The central point of the following sections is that these two sectors do not remain an
ordinary direct product once the Wess--Zumino coupling and the background flux
\(\mathcal F_{[2]}\) are taken into account. Their simultaneous gauging leads instead
to a \(2\)-group structure.

%%%%%%%%%%%%%%%%%%%%%%%%%%%%%%%%%%%%%%%%%%%%%%%%%%%%%%%%%%%%%%%%%%%%%%%%%%%%%%%%
\section{Gauging the compact and winding sectors}
\label{sec:gauging-symmetries}
We now couple the two generalized symmetry sectors of the compact toroidal theory to background gauge fields in order to gauge the global symmetry. The one-form gauge fields, $\mathcal B^r$, couple to the monopole symmetry generated by \eqref{eq:ungauged-shift-current} and a two-form background, $\widetilde{\mathcal B}^{\,r}$, to the winding symmetry \eqref{eq:winding-current}.\\

\noindent
As explained in \cite{Caro-Perez:2026pbu}, the standard gauging of the compact shift symmetry is obtained by promoting the global shift \eqref{eq:26} a local transformation. It is straightforward to verify that the action is gauge invariant under this transformation, provided one defines the covariant derivative
\begin{align}
    DX^r
    :={}&
    dX^r-\mathcal B^r.
    \label{eq:DX-def}
\end{align}
The ordinary \(0\)-form gauge transformations are
\begin{align}
    X^r
    &\longmapsto
    X^r+\varepsilon^r,
    \\
    \mathcal B^r
    &\longmapsto
    \mathcal B^r+d\varepsilon^r.
\end{align}
Therefore the $DX^r$ is gauge invariant. The gauging the shift symmetry is obtained from the replacement \(dX^r\mapsto DX^r\):
\begin{align}
    S_{M2}^{g}
    ={}&
    -\frac12
    \int_{\Sigma_3}
    DX^r\wedge\star DX_r
    +
    \frac q2
    \int_{\Sigma_3}
    \mathcal A_{[1]}\wedge
    \epsilon_{rs}DX^r\wedge DX^s
    +\notag
    \\
    {}&+
    S_{M2}^{nc+S^2}.
    \label{eq:shift-gauged-action}
\end{align}
Where by $S_{M2}^g$ we denote the gauged action. The term linear in \(\mathcal B^r\) is
\begin{align}
    \int_{\Sigma_3}
    \mathcal B^r\wedge J_r .
\end{align}
Thus \(\mathcal B^r\) is the background connection coupled to the improved monopole-like current \(J_r\). 
After gauging the shift symmetry, the gauge-invariant representative of the winding
current (the bianchi sector of 1-form of the embedding map) is
\begin{align}
    j^r_{\mathrm w}[\mathcal B]
    =
    DX^r.
\end{align}
If the winding symmetry were independent of the shift symmetry, one would introduce an ordinary two-form exact background  transformation as
\begin{align}
    \widetilde{\mathcal B}^{\,r}
    \longmapsto
    \widetilde{\mathcal B}^{\,r}
    +
    d\Lambda_{[1]}^{\,r},
\end{align}
and coupling it through
\begin{align}
    \frac12
    \int_{\Sigma_3}
    \widetilde{\mathcal B}^{\,r}\wedge DX_r.
\end{align}
in order to happen the 't~hoft anomaly as in \cite{Caro-Perez:2026pbu}.
However, because the Wess--Zumino cotribution the action, contains a monopole potential
\(\mathcal A_{[1]}\), the simultaneous gauging of the compact shift sector and the
winding sector requires an additional local term. We first write the most
useful boundary functional as
\begin{align}
    S_{M2}^{g}
    ={}&
    -\frac12
    \int_{\Sigma_3}
    DX^r\wedge\star DX_r
    +
    \frac q2
    \int_{\Sigma_3}
    \mathcal A_{[1]}\wedge
    \epsilon_{rs}DX^r\wedge DX^s
    +\notag
    \\
    +{}&
    S_{M2}^{nc+S^2}
    +
    \frac 12
    \int_{\Sigma_3}
    \widetilde{\mathcal B}^{\,r}\wedge DX_r
    +
    \frac q2
    \int_{\Sigma_3}
    \epsilon_{rs}
    \mathcal B^r\wedge\mathcal A_{[1]}\wedge DX^s .
    \label{eq:trial-gauged-action}
\end{align}
The last term is the local improvement forced by the Wess--Zumino coupling to the background field. 

\noindent
Let us first apply the ordinary product-type gauging transformations
\begin{align}
    X^r
    &\longmapsto
    X^r+\varepsilon^r,
    \\
    \mathcal B^r
    &\longmapsto
    \mathcal B^r+d\varepsilon^r,
    \\
    \widetilde{\mathcal B}^{\,r}
    &\longmapsto
    \widetilde{\mathcal B}^{\,r}
    +
    d\Lambda_{[1]}^{\,r}.
    \label{eq:ordinary-product-transformations}
\end{align}
Since \(DX^r\) is invariant, the kinetic term and the gauged Wess--Zumino term are
invariant. However, the new term in \eqref{eq:trial-gauged-action} varies under
the \(0\)-form gauge transformation as
\begin{align}
\delta_{\varepsilon}S_{M2}^{g}
    =
    \frac q2
    \int_{\Sigma_3}
    \epsilon_{rs}
    d\varepsilon^r\wedge\mathcal A_{[1]}\wedge DX^s .
    \label{eq:first-anomalous-term}
\end{align}
This is the first anomalous variation. It signals that the winding background cannot
transform independently if the Wess--Zumino coupling is present.

\noindent
The second anomalous variation comes from the ordinary \(1\)-form gauge
transformation of \(\widetilde{\mathcal B}^{\,r}\). One finds
\begin{align}
    \delta_{\Lambda}S_{M2}^{g}={}&
    \frac12
    \int_{\Sigma_3}
    d\Lambda_{[1]}^{\,r}\wedge DX_r.
\end{align}
Using $dDX_r=-d\mathcal B_r$, we obtain, up to a total derivative,
\begin{align}
    \delta_{\Lambda}S_{M2}^{g}
    ={}&
    -\frac12
    \int_{\Sigma_3}
    \Lambda_{[1]}^{\,r}\wedge d\mathcal B_r .
    \label{eq:second-anomalous-term}
\end{align}
Therefore, under the ordinary product-type transformations, the simultaneous gauging
produces two non-invariant terms:
\begin{align}
    \delta S_{M2}^{g}={}&
    \frac12
    \int_{\Sigma_3}\bigg(q
    \epsilon_{rs}
    d\varepsilon^r\wedge\mathcal A_{[1]}\wedge DX^s -\Lambda_{[1]}^{\,r}\wedge d\mathcal B_r \bigg).
    \label{eq:two-anomalous-terms-ordinary}
\end{align}
The first term is not a genuine 't Hooft anomaly. Rather, it means that the assumption that the two backgrounds transform independently is incorrect. We will see in the following that is removed by the \(2\)-group transformation law in analogy with the results found in \cite{Cordova:2018cvg}. The second term is the mixed 't Hooft anomalous variation that remains on the boundary and is cancelled by inflow.

\noindent
We now replace the ordinary transformation of \(\widetilde{\mathcal B}^{\,r}\) by the
\(2\)-group transformation
\begin{align}\label{eq:full-two-group-transformations}
    \widetilde{\mathcal B}^{\,r}
    \longmapsto
    \widetilde{\mathcal B}^{\,r}
    +
    d\Lambda_{[1]}^{\,r}
    +q
    \epsilon^{rs}
    d\varepsilon_s\wedge\mathcal A_{[1]}.
\end{align}
The extra term in \eqref{eq:ordinary-product-transformations} contributes to the
variation of
\begin{align}
    \frac12
    \int_{\Sigma_3}
    \widetilde{\mathcal B}^{\,r}\wedge DX_r
\end{align}
as
\begin{align}
    \delta_{\varepsilon}^{2\text{-}\mathrm{grp}}
    \left(
    \frac12
    \int_{\Sigma_3}
    \widetilde{\mathcal B}^{\,r}\wedge DX_r
    \right)
    =
    -\frac q2
    \int_{\Sigma_3}
    \epsilon_{rs}
    d\varepsilon^s\wedge\mathcal A_{[1]}\wedge DX^r .
\end{align}
This cancels exactly \eqref{eq:first-anomalous-term}. Thus, after imposing the
\(2\)-group transformation law, the only remaining variation of the boundary action is \eqref{eq:second-anomalous-term}. This is the mixed 't Hooft anomaly for the background fields.

\noindent
It is useful to package the \(2\)-group transformation into the shifted two-form background
\begin{align}
    \mathbb B^{\,r}:=
    \widetilde{\mathcal B}^{\,r}
    -q
    \epsilon^{rs}
    \mathcal B_s\wedge\mathcal A_{[1]}.
    \label{eq:two-group-Bbb-def}
\end{align}
Indeed, using
\begin{align}
    \delta\mathcal B_s=d\varepsilon_s,
    \qquad
    \delta\widetilde{\mathcal B}^{\,r}
    =
    d\Lambda_{[1]}^{\,r}
    +q
    \epsilon^{rs}
    d\varepsilon_s\wedge\mathcal A_{[1]},
\end{align}
one obtains
\begin{align}
    \delta\mathbb B^{\,r}
    =
    d\Lambda_{[1]}^{\,r}.
\end{align}
In terms of \(\mathbb B^{\,r}\), the boundary action becomes
\begin{align}
    \mathcal{S}_{M2}^{g}
    ={}&
    -\frac12
    \int_{\Sigma_3}
    DX^r\wedge\star DX_r
    +
    \frac q2
    \int_{\Sigma_3}
    \mathcal A_{[1]}\wedge
    \epsilon_{rs}DX^r\wedge DX^s
    +\notag
    \\
    {}&+
    S_{M2}^{{nc+S^2}}
    +
    \frac12
    \int_{\Sigma_3}
    \mathbb B^{\,r}\wedge DX_r.
    \label{eq:boundary-action-Bbb}
\end{align}
This form makes the \(2\)-group covariance manifest. Under the local shift transformation, \(DX^r\) and \(\mathbb B^{\,r}\) are invariant. Under the \(1\)-form gauge transformation and the boundary action varies precisely as in \eqref{eq:second-anomalous-term}.

\noindent
To cancel this remaining anomalous variation, we introduce a auxiliar four-manifold \(D_4\) with
\begin{align}
    \partial D_4=\Sigma_3,
\end{align}
and extend the background fields to \(D_4\). The required inflow term is
\begin{align}
    S_{\mathrm{inflow}}
    ={}&
    \frac12
    \int_{D_4}
    \mathbb B^{\,r}\wedge d\mathcal B_r .
    \label{eq:inflow-Bbb}
\end{align}
uniquely modified by the definition of $\mathbb{B}^r$. Then,
\begin{align}
    \delta_\Lambda S_{\mathrm{inflow}}
    =
    \frac12
    \int_{D_4}
    d\Lambda_{[1]}^{\,r}\wedge d\mathcal B_r
    =
    \frac12
    \int_{\Sigma_3}
    \Lambda_{[1]}^{\,r}\wedge d\mathcal B_r,
\end{align}
where Stokes' theorem was used. Hence
\begin{align}
    \delta_{\epsilon,\Lambda}
    \left(
    \mathcal{S}_{M2}^{g}
    +
    S_{\mathrm{inflow}}
    \right)
    =
    0.
\end{align}
Thus the boundary theory has a mixed 't Hooft anomaly, while the TQFT inflow term contain the oposites anomaly term cancellating it. The total M2-brane gauge-invariant functional is
\begin{align}
    \mathcal{S}_{M2}^{g}
    =
    {}&
    -\frac12
    \int_{\Sigma_3}
    DX^r\wedge\star DX_r
    +
    \frac q2
    \int_{\Sigma_3}
    \mathcal A_{[1]}\wedge
    \epsilon_{rs}DX^r\wedge DX^s
    +\notag
    \\
    +{}&
    S_{M2}^{nc+S^2}+
    \frac12
    \int_{\Sigma_3}
    \mathbb B^{\,r}\wedge DX_r
    +
    \frac12
    \int_{D_4}
    \mathbb B^{\,r}\wedge d\mathcal B_r .
    \label{eq:total-gauge-invariant-functional}
\end{align}
The corresponding \(2\)-group field strength is
\begin{align}
    H^r:={}&
    d\tilde{\mathcal{B}^r}+q\epsilon^{rs}\mathcal{B}_s\wedge \mathcal{F}_{[2]} .
    \label{eq:H-two-group}
\end{align}
It is invariant under \eqref{eq:full-two-group-transformations}. Moreover, since
\(d\mathcal F_{[2]}=0\), it satisfies the twisted Bianchi identity
\begin{align}
    dH^r
    =
    q\epsilon^{rs}
    d\mathcal B_s\wedge\mathcal F_{[2]}.
    \label{eq:H-bianchi-two-group}
\end{align}
Using \eqref{eq:calA-calF-def} this becomes
\begin{align}
    dH^r
    =q\epsilon^{rs}
    d\mathcal B_s\wedge(f_2\omega_1+f_3\omega_2).
\end{align}
This is the local differential signature of the \(2\)-group structure, the compact
\(0\)-form monopole sector and the winding \(1\)-form sector are not independent
background systems. Its simultaneous measurement is controlled by the monopole curvature on the sphere. See that the monopole curvature of the 2-torus does not contribute to in the 4-form $dH^r$.

%%%%%%%%%%%%%%%%%%%%%%%%%%%%%%%%%%%%%%%%%%%%%%%%%%%%%%%%%%%%%%%%%%%%%%%%%%%%%%%%
\section{Equations of motion and Page-like current}
\label{sec:eom-page-current}

We now derive the compact-sector equation of motion in the presence of the
\(2\)-group conecction fields. We use the action given by:
\begin{align}
   \mathcal{S}_{M2}^{g}={}&
    -\frac12
    \int_{\Sigma_3}
    DX^r\wedge\star DX_r
    +
    \frac q2
    \int_{\Sigma_3}
    \mathcal A_{[1]}\wedge
    \epsilon_{rs}DX^r\wedge DX^s
    +\notag
    \\
    {}&+
    \frac12
    \int_{\Sigma_3}
    \mathbb B^{\,r}\wedge DX_r
    +\frac12
    \int_{\Sigma_3}
    \mathbb B^{\,r}\wedge DX_r+
    S_{M2}^{nc+S^2}
    \label{eq:boundary-action-for-eom}
\end{align}
The background fields \(\mathcal B^r\), \(\widetilde{\mathcal B}^{\,r}\), and
\(\mathcal A_{[1]}\) are kept fixed when varying the embedding fields \(X^r\).
Therefore
\begin{align}
    \delta DX^r=d(\delta X^r).
\end{align}
The variation of \eqref{eq:boundary-action-for-eom} can be written in the form
\begin{align}
    \delta \mathcal{S}_{M2}^{g}
    =
    -\int_{\Sigma_3}
    d(\delta X^r)\wedge
    \left[
    \star DX_r
    +
    q\epsilon_{rs}\mathcal A_{[1]}\wedge DX^s
    -
    \frac12\mathbb B_r
    \right]=0
\end{align}
After integrating by parts, and assuming that \(\Sigma_3\) is closed, one obtains
\begin{align}
  \widetilde{J}_m^{\,r}=\star DX_r
    +
    q\epsilon_{rs}\mathcal A_{[1]}\wedge DX^s
    -
    \frac12\mathbb B_r, \quad  d\widetilde J_m^{\,r}=0.
    \label{eq:eom-current-conservation}
\end{align}
In terms of the \(2\)-group curvature \eqref{eq:H-two-group} we have:
\begin{align}
    d\star DX^r
    +
   q \epsilon^{rs}\mathcal F_{[2]}\wedge DX_s-
    \frac12 H^r
    +
    \frac{3q}{2}
    \epsilon^{rs}
    d\mathcal B_s\wedge\mathcal A_{[1]}=0,
    \label{eq:eom-H-form}
\end{align}
Thus the ungauged conservation law \(dJ_r=0\) is replaced, in the presence of the \(2\)-group backgrounds, by the dressed conservation law \(d\widetilde J_m^{\,r}=0\).

\noindent
Under the full \(2\)-group gauge transformations, the shifted two-form satisfies
\begin{align}
    \mathbb B^{\,r}
    \longmapsto
    \mathbb B^{\,r}+d\Lambda_{[1]}^{\,r},
\end{align}
while \(DX^r\) is invariant. Therefore
\begin{align}
    \widetilde J_m^{\,r}
    \longmapsto
    \widetilde J_m^{\,r}
    -\frac{1}{2}d\Lambda_{[1]}^{\,r}.
    \label{eq:Page-current-transformation}
\end{align}
Thus, for a closed two-cycle \(\mathcal N_2\subset\Sigma_3\), the integral
\begin{align}
    \int_{\mathcal N_2}
    \widetilde J_m^{\,r}
\end{align}
is invariant under small \(1\)-form gauge transformations. Under large
transformations, it is well-defined modulo the standard quantization of the
\(1\)-form gauge parameter. The associated operator is
\begin{align}
    U_\alpha^{(m),r}(\mathcal N_2)
    =
    \exp\!\bigg(
    i\alpha
    \int_{\mathcal N_2}
    \widetilde J_m^{\,r}
    \bigg).
\end{align}
The operator \(U_{\beta}^{(\mathrm w),r}(\mathcal C_1)\) is invariant under small \(0\)-form gauge transformations. Under large gauge transformations, the operator is well-defined modulo the standard quantization of the compact scalar \(X^r\) and of the background connection \(\mathcal B^r\), as in the standard coupling of compact fields to background gauge fields~\cite{Gaiotto:2014kfa,Sharpe:2015mja}. There is no inconsistency in allowing non-flat compact-shift backgrounds.
\begin{align}
    d\mathcal B^r\neq 0.
\end{align}
On the contrary, such backgrounds are precisely the ones that probe the full
\(2\)-group structure. What changes is that the gauge-invariant winding current
\begin{align}
    j^{r}_{\mathrm w}[\mathcal B]
    =
    DX^r
\end{align}
is no longer closed:
\begin{align}
    d j^{r}_{\mathrm w}[\mathcal B]
    =-d\mathcal B^r .
    \label{eq:winding-current-nonclosed}
\end{align}
Thus the winding operator
\begin{align}
    U_{\beta}^{(\mathrm w),r}(\mathcal C_1)
    =
    \exp\!\bigg(
    i\beta
    \int_{\mathcal C_1}
    DX^r
    \bigg)
\end{align}
is topological only in the flat sector
\begin{align}
    d\mathcal B^r=0.
\end{align}
For the non-flat \(\mathcal B^r\), the failure of
\(U_{\beta}^{(\mathrm w),r}(\mathcal C_1)\) to be topological is not a pathology of
the construction. Rather, it is the expected background-field signal of the fact that
the winding sector is not independent of the compact shift sector. This is precisely
the mechanism underlying a \(2\)-group symmetry, the background associated with the HFS$(1)$ is not an independent closed field once the background for the
HFS$(0)$ is turned on. Instead, its gauge-invariant curvature is modified
by the \(0\)-form background, and the corresponding Bianchi identity is twisted by
the Postnikov data of the \(2\)-group other relates cases
\cite{Benini:2018reh,Cordova:2018cvg}. Closely related realizations occur in
systems where winding symmetries are mixed with ordinary flavor symmetries by
Wess--Zumino type couplings \cite{Davighi:2024zjp}.

\noindent
In the present M2-brane compact sector, this structure is realized by the \(2\)-group curvature whose Bianchi identity is \eqref{eq:H-bianchi-two-group}.
Thus the curvature \(d\mathcal B^r\) of the compact-shift background is not an
inconsistency. It is precisely the background source that detects the Postnikov obstruction of the
\(2\)-group, as discussed in Section~\ref{sec:postnikov-exact-background}. The special case
\begin{align}
    d\mathcal B^r=0
\end{align}
will be discussed separately in Section~\ref{sec:flat-background-exact}. In that restricted sector the local differential signature of the \(2\)-group becomes
invisible, since
\begin{align}
    dH^r=0,
\end{align}
and the winding operator becomes topological again. This does not mean that the \(2\)-group is absent. It only means that flat compact-shift backgrounds do not probe
the Postnikov class.

\section{Postnikov class of M2-brane in \texorpdfstring{$AdS_5\times S_{1}^2\times S_2^2\times T^2$}{AdS5 x S1^2 x S2^2 x T2}}
\label{sec:postnikov-exact-background}
The previous sections show that the compact shift sector and the winding sector do
not define two independent background systems. Instead, they combine into a
\(2\)-group, this structure is described by a fibration of classifying
spaces
\begin{align}
    B^2U(1)_{\mathrm w}^{\,2}
    \longrightarrow
    B\mathbb G_{\mathrm{M2}}
    \longrightarrow
    BU(1)_{\mathrm m}^{\,2}.
    \label{eq:two-group-fibration-M2}
\end{align}
Here \(U(1)_{\mathrm m}^{\,2}\) is generated by the two compact shifts of \(X^r\), while \(U(1)_{\mathrm w}^{\,2}\) is the winding HFS$(1)$. The failure of this fibration to split is measured by a Postnikov class.

\noindent
In our case, the relevant \(0\)-form backgrounds are the \(U(1)\) connections
\(\mathcal B^r\), whose Chern classes are
\begin{align}
    c_1^{(r)}
    =
    \left[
    \frac{d\mathcal B^r}{2\pi}
    \right].
\end{align}
The monopole data entering the \(2\)-group is the diagonal class associated with
\(\mathcal F_{[2]}\), defined in \eqref{eq:calA-calF-def}. Equivalently, using the normalization of the spherical forms in \eqref{eq:normalization-volumen-forms-spheres},
\begin{align}
    u_{\mathcal F}
    :=
    \left[
    \frac{\mathcal F_{[2]}}{2\pi}
    \right].
\end{align}
Thus the universal Postnikov data is\footnote{This Postnikov datum can also be formulated via the \textit{Bockstein} map in $H^3(BU(1)^2_m\times S^2_1\times S^2_2,U(1))$, as in \cite{Benini:2018reh}. We will not use this description here.}
\begin{align}
[\beta^r]_{\mathrm{univ}}
    ={}&
    q
    \epsilon^{rs}
    c_1^{(s)}
    \smile
    u_{\mathcal F} \in H^4(BU(1)^2_m\times S^2_1\times S^2_2,\mathbb{Z}).
    \label{eq:postnikov-universal}
\end{align}
Where $c_1^{(s)}\smile u_{\mathcal F}$ is the  cup product \cite{Cordova:2018cvg}. In the flux normalization where
\begin{align}
    u_{\mathcal F}=f_2u_1+f_3u_2,
\end{align}
with $(f_2,f_3)\in \mathbb{Z}^2$ y que cumplan con \textit{AEC} \cite{Wulff:2016vqy}, this becomes
\begin{align}
    [\beta^r]_{\mathrm{univ}}
    =q
    \epsilon^{rs}
    c_1^{(s)}
    \smile
    (f_2u_1+f_3u_2).
    \label{eq:postnikov-universal-diagonal}
\end{align}
This is the integral representative of the obstruction to factorizing the compact \(0\)-form sector and the winding \(1\)-form sector. Equivalently, the obstruction is
controlled by the cup product between the Chern class of the compact monopole background and the diagonal monopole class induced by the quantized M-theory flux \cite{Witten:1996md}:
\begin{align}
    c_1^{(s)}
    \smile
    u_{\mathcal F}.
\end{align}

\noindent
Thus the coefficient entering the Postnikov class is the integral flux vector
\((qf_2,qf_3)\), which multiplies the monopole classes of the two spherical
factors. In this way, the obstruction is controlled by the mixed flux class
\(u_{\mathcal F}=f_2u_1+f_3u_2\) together with the winding integer \(q\).

\noindent
The differential representative of this class is precisely the twisted Bianchi
identity found above \eqref{eq:H-bianchi-two-group}. Therefore the local mixing of
the \(2\)-group backgrounds is the differential-form realization of the integral
Postnikov class \eqref{eq:postnikov-universal-diagonal}.

\noindent
Let us finally clarify the normalization of the coefficient. The parameter \(f\) in the local supergravity flux \eqref{eq:F4-ads5-s2-s2-t2} fixes the local curvature
scale of the solution. However, the integral Postnikov class is controlled by the
normalized M-theory flux integer. The \(C\)-field obeys the usual M-theory flux
quantization condition \cite{Witten:1996md}
\begin{align}
    \left[
    \frac{F_{[4]}}{(2\pi\ell_p)^3}
    \right]
    -
    \frac12\lambda
    \in
    H^4(M_{11},\mathbb Z)
    ~~~ \text{with} ~~~
    \lambda=\frac12p_1(T\mathcal{M}_{11}),
\end{align}
whete $p_1(T\mathcal{M}_{11})$ is the Pontryagin class. Suppressing the gravitational shift on
the cycles considered here, the normalized flux defines an integral class. Its mixed toroidal component pushes forward to
\begin{align}
    u_{\mathcal F}
    =
    \pi_{T^2\,*}
    \left[
    \frac{F_{[4]}}{(2\pi\ell_p)^3}
    \right]
    =
    f_2u_1+f_3u_2,
    \label{eq:uF-pushforward-short}
\end{align}
Therefore the coefficient entering the
Postnikov class is the integer \((qf_2,qf_3)\), not an arbitrary real parameter. 
Consequently, the \(2\)-group does not by itself impose the original M-theory flux
quantization. Rather, its Postnikov class is well-defined integrally because the
coefficient that enters it is the normalized \(F_{[4]}\)-flux integer inherited from the quantization of the \(C_{[3]}\)-field. In this sense, the \(2\)-group remembers the quantized
M-theory flux seen by the M2-brane through its Wess--Zumino coupling.

%%%%%%%%%%%%%%%%%%%%%%%%%%%%%%%%%%%%%%%%%%%%%%%%%%%%%%%%%%%%%%%%%%%%%%%%%%%%%%%%
%%%%%%%%%%%%%%%%%%%%%%%%%%%%%%%%%%%%%%%%%%%%%%%%%%%%%%%%%%%%%%%%%%%%%%%%%%%%%%%%
\section{Flat background sector}
\label{sec:flat-background-exact}

We finally comment on the special sector in which the monopole background is flat,
\begin{align}
    d\mathcal B^r=0.
\end{align}
Using the twisted Bianchi identity \eqref{eq:H-bianchi-two-group},
\begin{align}
    dH^r
    =q
    \epsilon^{rs}
    d\mathcal B_s\wedge\mathcal F_{[2]},
\end{align}
one immediately obtains
\begin{align}
    dH^r=0.
\end{align}
Thus, in this restricted background sector, the local differential signature of the
\(2\)-group is no longer visible. Equivalently, the gauge-invariant winding current
\begin{align}
    j^r_{\mathrm w}[\mathcal B]
    =
    DX^r
\end{align}
becomes closed again, since
\begin{align}
    dDX^r
    =
    -d\mathcal B^r
    =
    0.
\end{align}
Therefore the winding operator
\begin{align}
    U_{\beta}^{(\mathrm w),r}(\mathcal C_1)
    =
    \exp\!\bigg(
    i\beta
    \int_{\mathcal C_1}
    DX^r
    \bigg)
\end{align}
is topological in the flat sector. This does not mean that the \(2\)-group is absent. Rather, it means that flat monopole backgrounds do not probe the Postnikov obstruction. The obstruction becomes locally detectable only when non-flat backgrounds,
\(d \mathcal{B}^r\neq0\), are allowed. This is the standard behaviour of \(2\)-group backgrounds, the \(1\)-form background is not an
independent closed field once the \(0\)-form background has non-trivial curvature \cite{Cordova:2018cvg,Benini:2018reh}. Similar mechanisms occur in Wess--Zumino systems where winding symmetries mix with ordinary flavor symmetries \cite{Davighi:2024zjp}.

\noindent
It would be interesting to understand whether this Postnikov obstruction has dynamical consequences for the spectrum of the compactified M2-brane. In particular,
one may ask whether it induces selection rules among winding sectors, modifies the classification of admissible wrapped configurations, or constrains the central-charge
sector for an M2-brane with spatial wrapping on \(T^2_{\Sigma}\) as happens in \cite{Boulton:2002br,GarciaDelMoral:2018jye}. We leave quations on spectral aspects for a future work.
0

%%%%%%%%%%%%%%%%%%%%%%%%%%%%%%%%%%%%%%%%%%%%%%%%%%%%%%%%%%%%%%%%%%%%%%%%%%%%%%%%
\section{Conclusions}
\label{sec:conclusions-exact-background}
It has been demonstrated that when formulated in the  \(\mathrm{AdS}_5\times S^2_1\times S^2_2\times T^2\) background with fluxes, the M2-brane carries a non-trivial higher-group structure once the Wess--Zumino coupling to the eleven-dimensional three-form field is included.

\noindent
The two toroidal embedding coordinates give rise to a compact shift symmetry and to the winding sectors. In the absence of the flux-induced coupling, these two sectors would behave as independent HFS$(1)$. However, in the presence of the Wess—Zumino term, this factorization is obstructed.

\noindent
The obstruction becomes manifest when the monopole and winding currents are coupled to the background fields. A product-type gauging produces a non-covariant variation, which is removed only if the winding background associated to $T^2$ transforms non-trivially under the gauge transformation of the monopole $S^2_{1}\times S^2_{2}$ background. This is the characteristic signature of a \(2\)-group symmetry. The corresponding higher curvature satisfies a twisted Bianchi identity, showing that the winding background is sourced by the curvature of the compact monopole background together with the diagonal monopole flux.

\noindent
The Postnikov class of this \(2\)-group is controlled by the diagonal monopole class induced by the supergravity flux \eqref{eq: flux_quantization}. In this context, the coefficients entering in this class do not belong to the local supergravity parameter-vector \((f_1,f_2,f_3)\), but to the normalized integral flux vector \((qf_2,qf_3)\) obtained from the quantization of the M-theory \(F_{[4]}\)-flux and from the wrapping \(q\ne 0\) of the M2-brane arround the \(T^2\), that induces a nontrivial central charge on the M2-brane \cite{Martin:1997cb}. Thus, the \(2\)-group background system remembers the quantized mixed flux seen by the M2-brane through its Wess--Zumino coupling.

\noindent
We also describe, as a particular case, the flat compact-background sector. It corresponds to imposing \(d\mathcal B^r=0\). In this case, the local differential signature of the \(2\)-group becomes invisible, and the winding operator becomes topological again \cite{Caro-Perez:2026pbu}. This does not mean that the \(2\)-group is absent. Rather, flat backgrounds do not probe the Postnikov obstruction. The full structure is detected only when general, non-flat monopole backgrounds are allowed.

\noindent
These results provide a direct worldvolume realization of a \(2\)-group symmetry in the bosonic M2-brane. They also suggest a possible connection with the spectral properties of compactified supermembranes with non-trivial winding or central charge \cite{Boulton:2002br}.
In this construction, we have not considered the presence of monopoles over the $T^2$-sector of space-time $\mathcal{M}_{11}$ induced by an irreducible wrapping because we have studied a maximal (super)symmetric theory \cite{Wulff:2016vqy}.
Another natural direction for future work is to determine whether the Postnikov class induces selection rules among winding sectors, constrains admissible wrapped configurations, or modifies the central-charge sector of the compactified M2-brane in backgrounds with non-trivial \(C_{[3]}\) connection.

% We also described, in a particular case, the flat compact-background sector. When \(d\mathcal B^r=0\), the local differential signature of the \(2\)-group becomes invisible and the winding operator becomes topological again \cite{Caro-Perez:2026pbu}. This does not mean that the \(2\)-group is absent. Rather, flat backgrounds do not probe the Postnikov obstruction. The full structure is detected only when general, non-flat monopole backgrounds are allowed.

% \noindent
% These results provide a direct worldvolume realization of a \(2\)-group symmetry in the bosonic M2-brane. They also suggest a possible connection with the spectral properties of compactified supermembranes with non-trivial winding or central charge \cite{Boulton:2002br}.
% In this construction, we have not considered the presence of monopoles over $T^2$-sector of space-time $\mathcal{M}_{11}$ because we have studied a maximal (super)symmetric theory \cite{Wulff:2016vqy}.
% Other natural direction for a future work is to determine whether the Postnikov class induces selection rules among winding sectors, constrains admissible wrapped configurations, or modifies the central-charge sector of the compactified M2-brane in backgrounds with non-trivial \(C_{[3]}\) connection.

%%%%%%%%%%%%%%%%%%%%%%%%%%%%%%%%%%%%%%%%%%%%%%%%%%%%%%%%%%%%%%%%%%%%%%%%%%%%%%%%%%%%%%%%%%%%%%%%%%%%%%%%%%%%%%%%%%%%%%%
\section*{Acknowledgements}
FCP thanks to Physics area of the Chemistry Departament at U. Rioja, Spain and the Science Computation Research institute at the U. Rioja (SCRIUR), where part of this work was done and thanks to Instituto de Ciencias Exactas y Naturales (ICEN) at U. Arturo Prat, Iquique-Chile, where part of this work was done. FCP is supported by Doctorado nacional (ANID) 2023 Scholarship N$21230379$, project MATH-AMSUD 240048, and supported as graduate student in the “Doctorado en Física Mención Física-Matemática” Ph.D. program at the Universidad de Antofagasta. 
MPGM has been  partially supported by the PID2024-155685NB-C21 MCI Spanish Grants and by the University of La Rioja project REGI2025/41. AR  want to thank to MATH-AMSUD 240048  project and the Scientific Research Computing Institute of the University of La Rioja (SCRIUR), Spain. 
\appendix
\setcounter{equation}{0}
\renewcommand{\theequation}{A.\arabic{equation}}
%%%%%%%%%%%%%%%%%%%%%%%%%%%%%%%%%%%%%%%%%%%%%%%%%%%%%%
\bibliographystyle{elsarticle-num}
\bibliography{biblio}
\end{document}